\documentstyle[12pt,epsfig]{article}

\def\lesssim{\mathrel{\hbox{\rlap{\hbox{\lower4pt\hbox{$\sim$}}}\hbox{$<$}}}}
\def\gtrsim{\mathrel{\hbox{\rlap{\hbox{\lower4pt\hbox{$\sim$}}}\hbox{$>$}}}}

\newcommand{\ffrac}[2]
  {\left( \frac{#1}{#2} \right)}

\newcommand{\be}{\begin{equation}}
\newcommand{\ee}{\end{equation}}

 1
\font\elevenrm=cmr10 scaled\magstep 1
 1

\def\ref{\hang\noindent}

\begin{document}
\vspace*{1.8cm}
  \centerline{\bf TAU AIR-SHOWERS SIGNATURE }
  \centerline{\bf OF ULTRA HIGH ENERGY NEUTRINOS}
\vspace{1cm}
  \centerline{Daniele Fargion}
\vspace{1.4cm}
  \centerline{Physics Department and INFN }
  \centerline{\elevenrm Rome University 1,
     Pl.A.Moro 2, 00185, Rome, Italy}
\vspace{3cm}
\begin{abstract}
The discover of Ultra High Energy Neutrino of astrophysical
nature may be already reached. Indeed upward and horizontal $\tau$
Air-showers  emerging from the Earth crust or
 mountain chains offer the best and most powerful signal of Ultra High Energy  UHE
 neutrinos   $\nu_{\tau}$, $ \bar\nu_{\tau} $ and $ \bar{\nu}_e $ at PeV and
higher energy. The multiplicity in $\tau$ Air-showers secondary particles,
 $N_{opt} \simeq 10^{12} (E_{\tau} / PeV)$, $ N_{\gamma} (< E_{\gamma} >
\sim  10 \, MeV ) \simeq 10^8 (E_{\tau} / PeV) $ ,
 $N_{e^- e^+} \simeq 2 \cdot 10^7 ( E_{\tau}/PeV) $ ,
 $N_{\mu} \simeq 3 \cdot 10^5 ( E_{\tau}/PeV)^{0.85}$ make easy
 its discover. UHE  $\nu_{\tau}$ , $\nu_{\tau}$  following Super Kamiokande
 evidence of neutrino flavour mixing, ($\nu_{\mu}\leftrightarrow \nu_{\tau}$),
 should be as abundant as $\nu_{\mu}$, $ \bar\nu_{\mu}$.
 Also anti-neutrino electrons, $\bar{\nu}_e$, near the Glashow W
 resonance peak, $E_{\bar{\nu_e}} = M^2_W / 2m_e \simeq 6.3 \c 10^{15}\, eV$,
 may generate $\tau$ Air-showers .  Such  horizontal $\tau$ air-showers by $\nu_{\tau} N$ and
  UHE  $\bar\nu_{e} e$  at PeV emerging from mountain high chain might be the most power-full UHE neutrino
imprint. Upward UHE $ \nu_{\tau}- N $ interaction on Earth crust at
horizontal edge and from below, their consequent  UHE $\tau$
air-showers beaming toward high mountains should  flash $\gamma$,$\mu$,X and
Cherenkov  lights toward  detectors located on the top of the mountain or balloons.
 Such upward $\tau$ air-shower may hit also  nearby satellite
  flashing them by short,  hard, diluted $\gamma-$burst at the edge of
 Gamma Ray Observatory BATSE   threshold. We identify already these rarest gamma
  events with recent (1994) discovered upward Terrestrial Gamma Flashes  (TGF);
   we show their very probable  UHE $\tau$ -UHE $\nu_\tau$ origin.
    Partial TGF Galactic signature and known galactic and extra-galactic
source location are discovered within known 47 TGF events at low,
${ \simeq 2\cdot 10^{-3}}$ probability threshold.
\end{abstract}
\vspace{2.0cm}

\section{How  UHE $\bar{\nu}_e$, $\nu_\tau$, $\bar{\nu}_\tau$ are in-written by   Tau Air-showers}
Ultra high energy astrophysical neutrino (UHE$\nu$) from PeVs
($\gtrsim 10^{15}$ eV) up to  ($10^{18}$ eV) EeV and GZK cut off
energies ($ \gtrsim 10^{19}$ eV) might be traced  by $\tau$
induced air showers and by their millions to hundred billions
multiplicity in secondaries particles. Indeed astrophysical PeVs UHE
anti-neutrino electrons, $\bar{\nu}_e$, near the Glashow W
resonance peak, $E_{\bar{\nu_e}} = M^2_W / 2m_e \simeq 6.3 \cdot
10^{15}\, eV$, (dominant over expected UHE PeV atmospheric
neutrino signals), may be observable by their secondary horizontal
$\tau$ air showers originated by UHE chain reaction $\bar{\nu}_e +
e \rightarrow W^- \rightarrow \bar{\nu}_{\tau} + \tau^-$ inside
the concrete rock of a high mountain and their consequent escape
and  decay in air flight. Also UHE  $\nu_\tau$,
$\bar\nu_{\tau} $ at ($10^{16}$ - $10^{17}$eV) interacting with
nuclear matter ($\nu_\tau$ $N$) must be observable because of flavor
mixing $\nu_{\mu}\leftrightarrow \nu_{\tau}$ shown
by Superkamiokande data; indeed   huge astrophysical distances are
larger than the oscillation ones even for small mass differences
(below $\Delta m_{ij}^2 \sim 10^{-4}$ eV $^2$).
  Therefore UHE   $\nu_{\tau}$ and $\bar{\nu_{\tau}}$ may be converted and they may
 reach us from high energy galactic sources, as pulsars, Supernova
 remnants  or galactic micro-quasars and SGRs , as well as from powerful
 extra-galactic AGNs, QSRs or GRBs,  even at highest (GZK)
 energy because of the large galactic (Kpcs) and extreme cosmic
 (Mpcs) distances:

 \begin{equation}
 L_{\nu_{\mu} - \nu_{\tau}} = 4 \cdot 10^{-3} \,pc \left(
 \frac{E_{\nu}}{10^{16}\,eV} \right) \cdot \left( \frac{\Delta m_{ij}^2
 }{(10^{-2} \,eV)^2} \right)^{-1}
 \end{equation}

 These Tau air-showers are detectable in deep valleys or on front
 of large mountain chains as Alps , Rocky Mountains, Ande, (Fargion,Aiello,Conversano) $(1999)$.
The mountain  and  the air act as a fine tuning multi filter detector: as a screen of
undesirable  noisy horizontal UHECR showers (mainly electro-magnetic ones,
 Cherenkov photons, X,gamma and most of muons);
 as a dense calorimeter for UHE$\bar{\nu_{\tau}}$ nuclear events (three order of magnitude
 denser than air); as a distance meter target correlating $\tau$ birth
 place and its  horizontal  air-shower opening origination with
 the cosmic ray energy density; as a characteristic  anti neutrino detector
 by the extreme resonant cross section  $\bar{\nu_e} - e$
  and the consequent fine-tuned energy (few $PeV$) shower events;
   as a very unique  source of dense muon bundles from a mountain
   by main tau hadronic air-showering.\\
   The vertical up-ward tau air-showers (by small arrival
 nadir angle) occur preferentially at low energies nearly
 transparent to the Earth ($E_{\nu} \sim 10^{15} - 10^{16} $ eV).
 The  oblique $\tau$ air showers (whose arrival directions have large
 nadir angle), may be  related also to higher energy $\nu_{\tau}$, or
 $\bar{\nu_{\tau}}$ nuclear interactions
 ($E_{\bar{\nu_{\tau}}} \geq 10^{17} - 10^{19}$ eV). Indeed these horizontal - upward UHE
 $\nu_{\tau}$ cross a smaller fraction of the Earth volume and
 consequently they suffer less absorption toward the horizon.
 Moreover the consequent ultra-relativistic  ($E_{\bar{\nu_{\tau}}} \geq 10^{17} - 10^{19}$ eV)
  tau may travel in atmosphere for few or even hundred $Kms$
   with no absorption before the decay  to the detector located at few Kms distance.
    On the contrary  the horizontal gamma, electron pairs and
     muon showers by primary (down-ward nearly horizontal)
   UHECR proton  are severely suppressed   ($\geq 10^{-3}$) after
   crossing  ($\geq 2*10^{3}$)$gm.cm^{-2}$,or equivalent at one
   atmosphere,  ($\geq 16 Kms $) of horizontal atmosphere target.\\

These huge horizontal or upward air-shower signals being at least
million to billion times more abundant than the original and
unique UHE $\tau$  or UHE $\mu$ track in underground Km cube
detectors are much easier to be discovered with no ambiguity.
These high energy PeVs tau air-shower are mainly of astrophysical nature. Indeed
they cannot be produced by PeV atmospheric neutrino secondaries
born in atmospheric muon flavour and oscillating in tau state, because their
high PeV energy and their consequent large oscillation lengths are
much (hundred times) longer  than the  Earth  diameter.
We remind that on the contrary the expected long tracks
of upward muons  in underground km$^3$ detectors are mostly noisy
signals by TeVs to tens of TeVs muons secondaries generated by
atmospheric neutrinos born by common cosmic ray interactions in
upper atmosphere. The real (looked for) astrophysical signals are
 upward PeV neutrino muons which are, unfortunately, suppressed by
Earth opacity. This make cube Km detector , in our opinion,
less favorite than  our Tau up-ward air-shower detection.
 Let us remind that upward $\nu_{\tau}$  at same energies
as noted by different authors are less (an order of magnitude) suppressed.
Present $\tau$ air shower is analogous to the  well-known
Learned and Pakwasa $(1995)$ "double bang" in underground neutrino
detectors. The novelty of the present "one bang in" (the rock, the mountain, the Earth) -
"one bang out" (the air) lays in the self-triggered explosive
nature of $\tau$ decay in flight and its consequent huge amplified
air shower signal  at a characteristic few Kms distance.

\section{The peculiar  UHE $\bar{\nu}_e$, $\nu_\tau$, $\bar{\nu}_\tau$  and $\tau$ interactions }

 Moreover the expected $ \nu_{\tau} $ signals, by their
secondary tau tracks at highest cosmic ray energy window $1.7\cdot
10^{21} \,eV > E_{\tau} > 1.6\cdot 10^{17} \, eV $, must exceed
the corresponding $ \nu_{\mu} $ (or muonic) ones, making UHE $
\nu_{\tau} $ above $0.1$ EeV the most probable UHE signal. Indeed,
the Lorentz-boosted tau range length grows (linearly) above muon
range, for $ E_{\tau} \geq 1.6 \cdot 10^8 GeV $; (see Fig (1)
eq.3): the tau track reaches its maxima extension, bounded not by
bremsstrahlung radiation length nor by pair production (eq. 2),
 but by growing nuclear (mainly photo-nuclear) and mainly ,later, by  electro-weak
interactions (eq. 4), $ R_{\tau_{\max}} \simeq 191\;Km$, at energy
$ E_{\tau} \simeq 3.8\cdot 10^9\;GeV$ in water.

\begin{figure}[t]
\begin{center}
 \includegraphics[width=0.8\textwidth,  bb=173 144 625 533] {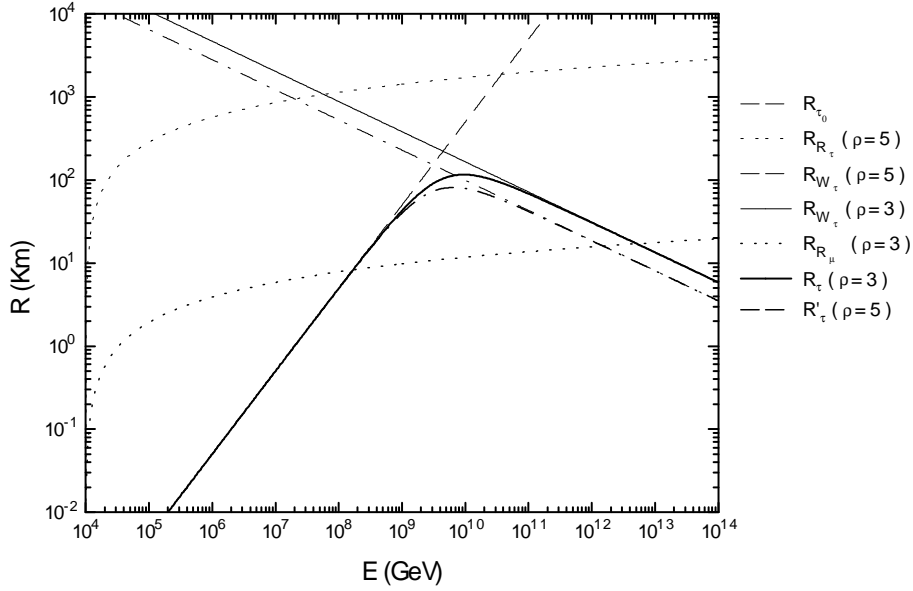}
\end{center}
  \caption{The tau ranges as a function of the tau energy respectively for
    tau lifetime (dashed line) $R_{\tau_o}$, for over-estimated tau radiation
    range $R_{R_\tau}$ ,(short dashed line above) and tau
    electro-weak interaction range $R_{W_\tau}$, for two densities
    $\rho_r$ (long dashed lines, continuous) and their combined range
    $R_{\tau}$. Below the corresponding radiation range
    $R_\mu$ for muons (dotted line).}
\label{fig:boxed_graphic 1}
\end{figure}

\begin{equation}\label{6}
 R_{R_{\tau}} \cong 1033 \; Km
\left(\frac{\rho_r}{5}\right)^{-1} \cdot
 \left\{\, 1 \,+\, \frac{\ln\left[\left(\frac{E_{\tau}}{10^8 \,
\mathrm{GeV}}\right)\left(\frac{E_{\tau}^{\min}}{10^4 \,
\mathrm{GeV}}\right)^{-1}\right]}{(\ln \, 10^4 )}\right\} .
\end{equation}



\begin{equation}\label{7}
R_{\tau_o} = c \tau_{\tau} \gamma_{\tau} = 5 \, \mathrm{Km} \,
\left(\frac{E_{\tau}}{10^8 \, \mathrm{GeV}}\right) \; .
\end{equation}

\begin{equation}\label{9}
 R_{W_{\tau}} = \frac{1}{\sigma N_A \rho_r} \simeq
\frac{2.6\cdot 10^3 \, \mathrm{Km}}{\rho_r} \,
\left(\frac{E_{\tau}}{10^8\, \mathrm{GeV}}\right)^{-0.363} \; .
\end{equation}

It should be noticed that the radiative $\tau$ length estimated
above has been considered for bremsstrahlung radiation length
only. Pair production energy loss is more restrictive in the final
$R_{R_\tau}$ length (by an approximate  factor
$\frac{m_\tau}{m_\mu}$) as well as the growing photo-nuclear
interactions at highest ( tens $EeV$) energies. However the very
dominant electro-weak interactions at these energies are already
suppressing the $\tau$ growth and the combined interaction length
are slightly less, but almost comparable to the one shown in
figure above.

 At the peak maxima the tau range is nearly $10$-$20$
times longer than the corresponding muon range (at the same
energy) implying, for comparable fluxes, a ratio $10$ times larger
in $ \nu_{\tau} $ over $ \nu_{\mu} $ detection probability. This
dominance, may lead to  a few rare spectacular event a year (if
flavor mixing occurs) preferentially in horizontal plane in
underground $Km^3$ detectors. The Earth opacity at those UHE
regimes at large nadir angles (nearly horizontal, few degree
upward direction) is exponentially different for UHE muons respect
to tau  at GZK  energies (corresponding to hundreds Kms UHE Tau
lengths), making the muon/tau flux ratio of such lengths severely
suppressed. Unfortunately there are not yet in underground
detector such possible tests; maybe in new generation  horizontal
slice detectors of  tens Km sizes the UHE GZK $\tau$  may
dominate by such lengths and huge showering. The ratio among
these $ \nu_{\tau} $ over $ \nu_{\mu} $ tens Kms signals is
exponentially high $> \exp(10)$.

\section{The  key role of UHE $\nu$,   interactions with the
relic $\bar{\nu}$ at Z resonance in hot dark haloes}

In the frame-work of UHE neutrino astrophysics it is important to
remind the very peculiar role that heaviest light neutrino (most
probably of tau nature) may play in the puzzling  GZK problem in
modern high energy astrophysics.Indeed let us remind here a
possible additional (and parental) role of UHE  $\nu$ above GZK
discussed in this article. UHE $ \nu_{\tau} $ at energies near or
above GZK energies are transparent to BBR cosmic photons
(contrary to common nucleons, nuclei, photons UHECR); therefore
neutrinos may easily reach us from far cosmological distances.
Then the puzzle of UHECR above GZK cut off may be solved assuming
that neutrinos (possibly of heaviest muon-tau nature) share a
light mass of few eV , in the frame-work of Hot Dark Matter halos
clustered around galaxies or local group. Such light neutrinos
may form a huge hidden dark calorimeter  able to beam dump UHE
$\nu$ via $Z$ (s-channel), via  virtual $W$ ($t$ channel)  or $W$
and $Z$ pair productions. The latter cross-sections are less
efficient than the  $Z$ (s-channel) but are not fine tuned or
restricted on a (very light, nearly half $eV$) relic neutrino
mass. The corresponding cross sections for such $\nu$ $\nu$
interactions are shown in Fig. 2; their secondaries may be final
UHE anti-protons (or anti-neutrons) or UHE protons (or neutrons)
(Fargion,Mele,Salis 1997-1999) responsible of final observed
UHECR above GZK cut off. The interaction efficiency by relic
light neutrinos via UHE $\nu$ at GZK cut off is thousands times
larger than UHE $\nu$ interactions on Earth atmosphere and/or
direct UHECR (nucleons,nuclei) propagation above GZK distances.
Therefore light neutrino mass may explain both hot dark matter
and UHECR above GZK (as well as their recent arrival clustering
in triplets or doublets). If this solution is correct we are
already testing both the neutrino mass  the relic neutrino
density and the $UHE$ neutrino flux. Just to underline the $\nu$
mass roles of a few $eV$ in modern high energy astrophysics  we
remind also the important case of a SN
 ejected neutrino burst (at $MeVs$ energies) arriving slowed by its mass relativistic
flight and delayed  respect to the arrival of the massless prompt super-nova
gravitational burst. The expected time delay between the massive neutrino with the massless
graviton wave burst, offer an additional test to the elusive neutrino mass detection:\\
 $ \Delta t\sim 50$ sec $ \ffrac{E_{\nu}}{5\,MeV}^{-2}
\ffrac{m_{\nu}}{5\,eV}^{2} \ffrac{L}{Mpc}$. (Fargion 1981).

\begin{figure}[bt]
\begin{center}
\includegraphics[width=0.8\textwidth]{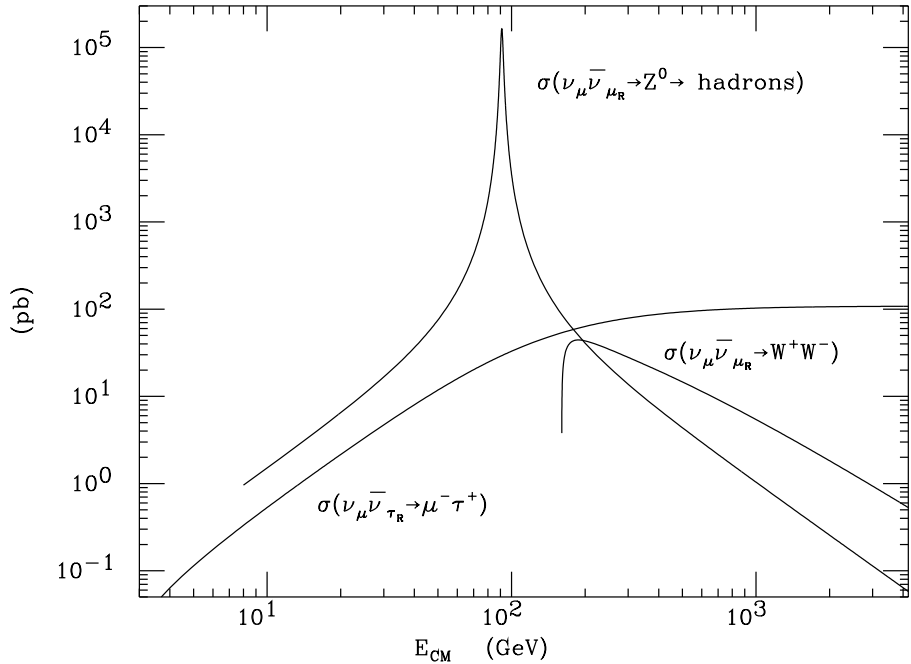}
\end{center}
  \caption{The total cross sections for the UHE $ \nu\bar{\nu}$ indicated processes
as function of the center of mass energy (Fargion,Mele,Salis 1999)}
\label{fig:boxed_graphic 2}
\end{figure}

%
\subsection{ Horizontal  $\tau$ air-shower born on front of mountain chain:  UHE  PeVs
neutrino trace}
 Therefore UHE Tau $ E_{\tau} \geq \ 10^5 GeV - 5 \cdot 10^7 GeV  $
air-shower in front of high mountains chains will be easily induce
peculiar horizontal UHE $\tau$ (Fargion, Aiello, Conversano 1999).
 Energies above will be probably missed.
 An hybrid detector (gamma/optical air-shower array)
 would get precise signal and arrival direction.
Because of the different neutrino interactions with energy
 it will be possible to estimate, by stereoscopic,
directional and time structure signature, the spatial air-shower
origination in air, the primary tau distance decay from the
mountain (tens  or a hundred of meters for fine tuned PeVs UHE $ \bar{\nu}_e $ and
 meters up to few Kms for  UHE $\nu_\tau$,$ \bar\nu_{\tau}$ at
 wider energy window $ E_{\tau} \geq \ 10^5 GeV - 5 \cdot 10^7 GeV $.
  Additional energy calibration may be derived sampling shower intensities.\\
   Hundreds of array (scintillator,Cherenkov) detectors
 in deep wide valley  horizontally oriented would be necessary to get tens
$\tau$ air--showers events a year. Screening by undesirable lateral
or downward noisy cosmic rays or natural radiation is achieved by
directional and time clustering filter; the induced
$\bar{\nu}_e e \rightarrow \tau$ air shower even in absence of
$\nu_{\mu} \leftrightarrow \nu_{\tau}$ oscillation should be well
identified and detectable. The atmospheric $\bar{\nu}_e$ ones (secondary by common UHECR)
 are suppressed by a  severe power law . Its unique $\bar{\nu}_e$ origin is
marked by the peaked  W resonance, and by the small mountain
$\bar{\nu}_e$ opacity and its high neutrino cross-section. Its
identity is marked by the expected fine tuned PeV energy at W peak.\\
 More copious ($> 5$ times more) events by PeV up
to tens PeV  charged current $\nu_{\tau} N$ interaction occur following
Super Kamiokande  flavour mixing discover.\\

  \subsection{ Upward $\tau$ PeV-EeV air-showers flashing
  to high  mountains, balloons and satellites}

 It will be also possible to observe UHE $\nu_{\tau}$,
by  the upward tau air-shower arriving from tens or  hundred
Kilometers away (near horizontal edges) from high mountains, high
balloon and satellites; such UHE tau created within a wide (tens
thousands to millions square km$^2$ wide and hundred meter UHE Tau
depth in Earth crust) target would discover only UHE
$\nu_\tau$,$ \bar\nu_{\tau}$ neutrinos at PeV up to EeV energies
and above, just within the mysterious GZK frontiers. The discover
will need capable gamma, optical and mainly muon bundle detectors
within present technology .\\
 From the same highest mountains, balloons and near orbit satellite, looking more
downward toward the Earth  it is possible to discover more
frequent but lower energetic astrophysical $\simeq$ PeV - tens PeV
 neutrinos still nearly transparent to the Earth volume (Gandhi et al. 1998), (see Fig.3).\\
\begin{figure}[bt]
\begin{center}
\includegraphics[width=0.8\textwidth]
{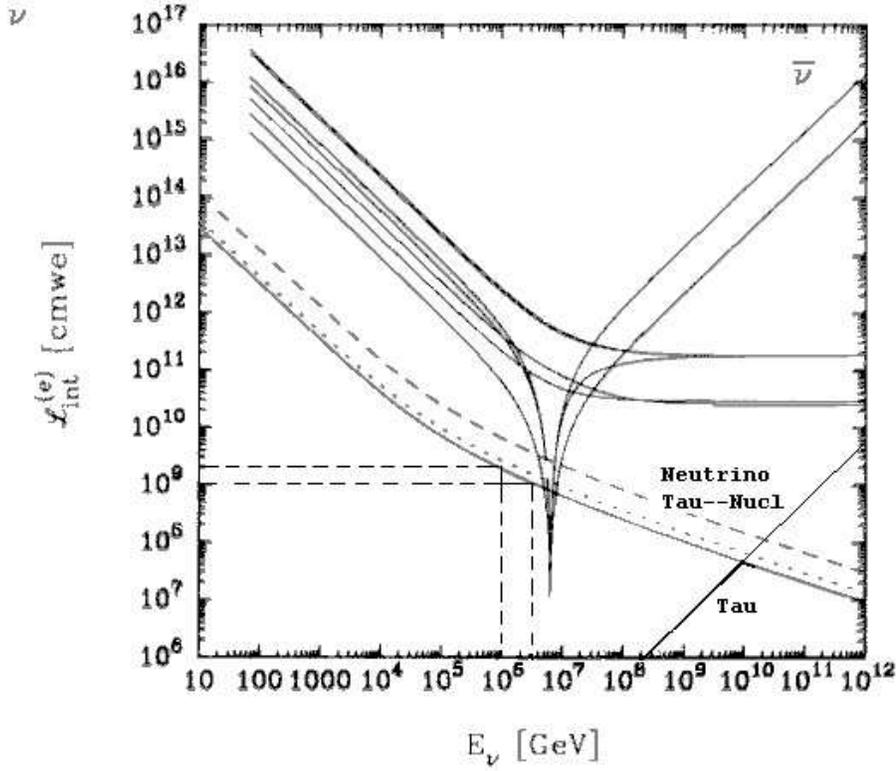}
\end{center}
  \caption {The Gandhi et all (1998) UHE neutrino
    ranges as a function of UHE neutrino energy in Earth with
    overlapping the resonant
    $\bar{\nu}_e e$, $\nu_\tau N$ interactions;
    below  in the corner the UHE $\tau$ range, as in Fig 1,
     at the  same energies in matter (water).\newline}
    \label{fig:boxed_graphic 5}
\end{figure}
The UHE neutrinos $\bar{\nu_e}$,${\nu}_{\mu}$ $\bar{\nu}_{\mu}$ are default and
expected UHECR ( $\gtrsim 10^{16}$ eV) secondary products near AGN
or micro-quasars by common photo-pion decay relics by optical
photons nearby the source (PSRs, AGNs) ($p + \gamma \rightarrow n
+ \pi^+, \pi^+ \rightarrow \mu^+ \nu_{\mu}, \mu^+ \rightarrow e^+
\nu_e \bar{\nu}_{\mu} $), or by proton proton scattering in
galactic interstellar matter. The maximal observational distances
from mountains, balloons or satellites, may
 reach $\sim$ 110 Km $(h/Km)^{\frac{1}{2}}$ toward the horizon,
 corresponding to a UHE $\tau$ energy $\sim 2 \cdot 10^{18}$ eV $(h/ Km)^\frac{1}{2}$.
 Therefore we propose to consider such upward shower nearly horizontal detection
 from high mountains to test this highest $\nu_{\tau}
 \bar{\nu_{\tau}}$ energy window almost opaque to Glashow UHE
 $\bar{\nu_e}$ fluxes just comparable to the GZK cut off energies.

The expected downward muon number of events $ N_{ev} (\bar{\nu}_e
e\to\bar{\nu}_{\mu} \mu)$ in the resonant energy range, in Km$^3$,
[Table 7,The Gandhi et all,1998] 
was found to be $N_{ev} = 6$ a year. One expect a comparable
number of reactions $ (\bar{\nu}_e e\to \bar{\nu}_{\tau} \tau) $.
However the  presence of primordial $\nu_{\tau}, \,
\bar{\nu_{\tau}}$ by flavor mixing and $\nu_{\tau}, \,
\bar{\nu_{\tau}} N$ charged current interactions lead to a factor
5 larger rate, $ N_{ev} = 29$ event/year.
\\ If one imagines a gamma/optical detector at 5 km far in front
of a chain mountain  as the Alps Argentier valley (size 10 km, height 1 km) one
finds a $\tau$ air shower volume observable within a narrow
 beamed cone (Moliere radius $\sim 80$ m / distance $\sim 5$ Km):
 ($\Delta \theta \sim    1^o$, $\Delta \Omega \sim 2 \cdot 10^{-5}$) and
 an effective volume $V_{eff} \simeq 9 \cdot 10^{-5} $
 Km$^3$ for each observational detector.
  Each single detector is comparable to
 roughly twice a Super Kamiokande detector.
  Following  common AGN - SS91 model [The Gandhi et all,1998] 
  we foresee a total event rate of:
  (6) ($\bar{{\nu}_e} e$) + (29) ($\nu_{\tau} N$) = 35 UHE $\nu_{\tau}$
  event/year/Km$^3$.
 At energies above 3 PeV we may expect a total rate of N$_{ev}$ $\sim$
 158 event/year in this Alps Argentiere mountains valley and
 nearly 3.2 $\cdot 10^{-3}$ event/year for each m$^2$ size detector.
 In a first approximation, neglecting Earth opacity,
  it is possible to show that the Earth volume observable from the top of
a mountain at height $h$, due to UHE $\tau$ at 3 PeV crossing from
below, is approximately V $\approx 5 \cdot 10^4$ Km$^3$
$\ffrac{h}{Km} \ffrac{E_{\tau}}{3\,PeV}$. These upward shower would
hit the top of the mountain. For the same $\tau$ air shower
beaming ($\Delta \theta \sim 1^o$, $\Delta \Omega \sim 2 \cdot
10^{-5}$) we derive now an effective volume $\sim$ 1 Km$^3$.
Therefore a detector open at $2 \pi$ angle on a top of a 2 Km
height mountain may observe nearly an  event every two month from
below the Earth. The gamma signal above few MeV would be
(depending on arrival nadir angle) between  $3\cdot 10^{-2}$
cm$^{-2}$ (for small nadir angle) to $10^{-5}$ cm$^{-2}$ at far
distance within 3 PeV energies. A contemporaneous (microsecond)
optical flash ($\gtrsim 300 \div 0.1 \, cm^{-2}$) must occur.
Keeping care of the Earth opacity, at large nadir angle ($\gtrsim
{60}^0$) where an average Earth density may be assumed ($< \rho >
\sim 5$) the transmission probability and creation of upward UHE
$\tau$ is approximately

\begin{equation}
P(\theta,\, E_{\nu}) = e^{\frac{-2R_{Earth} \cos
\theta}{R_{\nu_{\tau}}(E_{\nu})}} (1 - e^{-
\frac{R_{\tau}(E_{\tau})}{R_{\nu_{\tau}}(E_{\nu})}}) \, .
\end{equation}

This value, at PeV is within a fraction of a million(${\theta}{\approx}{60}{^0}$)
to a tenth of thousands (${\theta}{\approx}{90}{^0}$). The corresponding
angular integral effective volume observable from a high mountain
(or balloon) at height $h$ (assuming a final target terrestrial
density $\rho = 3$)  is:

\begin{footnotesize}
\begin{equation}
  V_{eff} \approx 0.3 \, Km^3 \ffrac{\rho}{3}\ffrac{h}{Km} e^{-
  \ffrac{E}{3\,PeV}}
  \ffrac{E}{3\,PeV}^{1.363}
\end{equation}
\end{footnotesize}

 A  popular "blazar" neutrino flux model (like Stecker Salomon, Berezinsky ones)
  normalized within a  flat spectra (at an
energy fluence  $\simeq 2 \,10^{3} eV \ cm^{-2}$) is leading,
above 3 PeV, to $\sim$ 10 UHE $\nu_{\tau}$ upward event/Km$^3$
year. Therefore we must expect an average upward effective event
rate observed on a top of a mountain (h $\sim 2\,Km$) (Fig. 4):

\begin{footnotesize}
\begin{equation}
  N_{eff} \simeq 8 \, \frac{\mathrm{events}}{\mathrm{year}} \ffrac{\rho}{3} \ffrac{h}{2 \,Km}
e^{- \ffrac{E}{3\,PeV}} \ffrac{E}{3\,PeV}^{1.363}
\end{equation}
\end{footnotesize}

This rate is quite large and one expected $\tau$ air air-shower
signal (gamma burst at energies $\gtrsim 10 \, MeV$) should be
$\phi_{\gamma} \simeq 10^{-4} \div 10^{-5}$ cm$^{-2}$, while the
gamma flux at ($\sim 10^5 \, eV$) or lower energies (from electron
pair bremsstrahlung) may be two order of magnitude larger. The
optical Cherenkov flux is large $\Phi_{opt} \approx 1$ cm$^{-2}$.

\section{Upward $\tau$ Air Shower in Terrestrial Gamma Flash: first evidences of UHE neutrinos}
The tau upward air showers born in a narrow energy
 window, $10^{15}$ eV $ \lesssim E_{\nu} \lesssim  5 \cdot 10^{16} $
 eV (Fig.3) may penetrate high altitude leaving rare beamed upward gamma
 shower bursts whose sharp ($\sim $ hundreds $\mu$sec because of the hundred kms high
 altitude shower distances) time
 structure and whose hard ($\gtrsim 10^{5}$eV) spectra may hit near
 terrestrial satellites.
We claim (Fargion 2000) that such gamma upward events originated
by tau air
 showers produce gamma bursts at the edge of GRO-BATSE sensitivity
 threshold. In particular we argue that very probably such upward
 gamma events have been already detected since April 1991 as
 serendipitous sharp ($\lesssim 10^{-3}$ sec) and hard ($\gtrsim
 10^5$ eV) BATSE gamma triggers originated from the Earth and
 named consequently as Terrestrial Gamma Flashes (TGF).

 \subsection{Time asymmetry in down-ward and up-ward Tau Air-shower }

Upward and Downward air-showers {\bf are not symmetric} event at
all because the different atmosphere densities at sea level and
high altitude. Indeed at sea level $\tau$ air-shower holds just a
$\mu$seconds. But at high level $\tau$ decays to produce
millisecond showers. \\  The arrival time of $\gamma$ air-shower
(bremsstrahlung photons) is ruled by the last atmosphere distances
where the   gamma emission has been originated (while being
nearly  unabsorbed). The mean energy deposition profile in air
shower  is given by a common gamma distribution:

     \begin{equation}
     \frac{dE}{dt} = \frac{E_0 b (bt)^{a-1} e^{-bt}}{\Gamma (a)}
     \end{equation}

   where the a-dimensional shower depth distance $t =
      \frac{x}{X_0}$ and the a-dimensional energy $y =
      \frac{E}{E_c}$, are well known variables. The characteristic
      critical energy $E_c$ (see B. Rossi or Longair text book) is, in air,
      around 100 MeV value. The air shower maxima occurs at an
      a-dimensional depth $t_{max} = \frac{a-1}{b}$, while the characteristic
      shower distance $X_s \equiv \frac{X_0}{b}$, being $b\simeq 0.5$, is
      $X_s \simeq 2X_0$; (note that
       $t_{max} = \frac{a-1}{b} \simeq (\ln y + \frac{1}{2}$)
        defines $a$ for a photon-induced cascades).
      Naturally the radiation length $X_0$ is the same for upward
      and downward air showers. However the corresponding length
      distances are very different because the different altitudes (sea level and
      high altitude) where the shower takes place,
       have extremely different densities. The air
      density decreases, respect to the sea level height, with
      altitude z as $\rho = \rho_0 e^{-\frac{z}{h_0}}$, $h_0 \simeq  8.55$ $Km$.
       If one considers the sea level case $X_0 = 36.6$
      g/cm$^2$ and the radiation distance is $X_0 = 304.2$ m,
      the shower length is $X_s \simeq 608.4$ m and the
      corresponding shower scale time is (as it is well known and as it is
      observed in common downward  air showers) $t_s = \frac{X_s}{c} \simeq 2 \mu
      s$.    If now we consider upward $\tau$ air shower arising on the top
       atmosphere altitude, than the same $X_0 = 36.6$ g/cm$^2$
        corresponds, in a more diluted upward atmosphere to distances
        $X'_0\simeq 22$ km and in a first approximation to a shower
        length $X'_s \simeq 44$ km leading to $t'_s \simeq 75 \, t_s \sim
        0.15$ ms. Additional time dilution  must be considered for the
        arrival nadir angle $\theta$: $t'_s \sim \frac{0.2}{\cos
          \theta}$ $ms$ and for geo-magnetic Larmor precession of relativistic
           electrons. More precisely, the $\tau$ air shower timing is
           related to the total distance from the earliest atmosphere
           last scattering ($X_0 \sim 36.6$ g/cm$^2$ and $h\simeq 22$
           km) up to the BATSE satellite height ($\sim 500$ km).
           To summarize the maximal upward $\tau$ air shower extend up to $t'_s
           \sim \frac{500}{c \, \cos \theta}$ km $ > 2$ ms. \\
           The exponential density decay in upper
           atmosphere makes most of the bremsstrahlung radiation
           generated at lowest (tens of kms altitudes) implying a fast raise of
           the gamma flash within a few tens of millisecond, (as the
           observed TGF ones) even if the gamma signal must
            also extend up to few millisecond
           times as indeed observed in TGFs. Different Tau Air-shower
           Channel (as well known in particle data text) and
           their consequent bifurcation may lead to rapid
           TGF millisecond timing modulations as the  observed ones.

   \subsection{ Why the Terrestrial Gamma Flash  are Tau Air-Showers}

The visible Earth surface from a satellite, like BATSE, at height
$h \sim 400$ Km and the consequent effective volume for UHE
$\nu_{\tau} N$ PeVs interaction and $\tau$ air shower beamed
within $\Delta \Omega \sim 2 \cdot 10^{-5} rad^2$ is: (note
$<\rho> \simeq 1.6$ because 70 \% of the Earth is covered by seas)
 $ V_{eff} = V_{TOT} \Delta \Omega \simeq 60 \, Km^3.$
The effective volume and the event rate should be reduced, at
large nadir angle ($\theta > 60^o$) by the atmosphere depth and
opacity (for a given $E_{\tau}$ energy). Therefore the observable
volume may be reduced approximately to within 15 Km$^3$ values and
the expected UHE PeV event rate is
\begin{equation}
  N_{ev} \sim 150 \cdot e^{-
  \ffrac{E_\tau}{3\,PeV}}
  \ffrac{E_\tau}{3\,PeV}^{1.363}
  \ffrac{h}{400 Km}\qquad
  \frac{\mathrm{events}}{\mathrm{year}}
\end{equation}

The TGF signals would be mainly $\gamma$ at flux $10^{-2}$
cm$^{-2}$ at X hundred keV energies. The observed TGF rate is
lower than the expected one (eq.~7) by nearly an order of magnitude, and this suggests higher
$E_{\nu}$ energies (to overcome BATSE threshold) and consequently
small additional probability suppression fitting the observed TGF
events rate. However since
 1994 (Fishman et al.) TGF understanding of presently known 75
 records over nearly eight thousand BATSE triggers is based on an
 unexpected and mysterious high latitude lightening of geophysical
 nature (the so called "Sprites" or "Blue Jets"). We do not
 believe in that interpretation.
 We notice that among the 75 records only 47 are published in
 their details, while 28 TGF are still unpublished. Their data
 release is therefore urgent and critical.
 While Blue Jets might be in principle triggered by upward tau
 air showers in the atmosphere (a giant "Wilson" room) we believe they are
 not themselves source of TGF. In particular their observed
 characteristic propagation velocity ($\lesssim 100$ Km/s) from
 distances $\sim$ 500 Km, disagree with short TGF millisecond
 timing and would favor a characteristic TGF time of few seconds. \\
  Moreover TGF data strongly dis-favor by its hard spectra
 the terrestrial Sprites connection. On the contrary the expected
 UHE tau upward air showers lead to a gamma burst flux,
 spectra, and fine time structure fluence in agreement with the
 observed TGF ones and in agreement with the expected
 flux models.
 The correlations of these clustered TGFs directions
 toward \\ (1) well known and maximal powerful galactic and
 extra-galactic sources either at TeV, GeV-MeV, X band ,(2) recent
 first anisotropy discovered on UHECR at EeV by AGASA, (see Fig.4,
 from Hayashida et al. 1999) (3) the
 Milky Way Galactic Plane (Fig.~4),  support
 and make compelling the TGF  identification as secondary gamma
 burst tail of UHE $\tau$ induced upward air shower.
   The present TGF-$\tau$ air-shower identification could not
    be produced by UHE $\bar{\nu}_e$ charged current
   resonant event at ($E_{\bar{\nu_{e}}} = M^2_W / 2m_e = 6.3 \cdot 10^{15}$
   eV), because of the severe Earth opacity for such resonant
   $\bar{\nu}_e$, and therefore it stand for the UHE $\nu_{\tau}
 \bar{\nu_{\tau}}$ existence. Consequently it gives support to the
 Superkamiokande evidences for $\nu_{\mu}\leftrightarrow \nu_{\tau}$
  flavor mixing from far PSRs or AGNs sources toward the Earth.
     At the present the very probable $\nu_{\tau}
     \bar{\nu_{\tau}}$ source of TGFs and their probable partial galactic location
     infer a first lower bound on
     $\Delta_{m_{\nu_{\mu} \nu_{\tau}}}$ ($L < 4$ Kpc, $\Delta_{m_{\nu_{\mu} \nu_{\tau}}}
      > 10^{-8}$ eV$^2$)
     and it offers a first direct test of
     the same existence of the last evanescent (hardly observed only recently),
     fundamental neutral lepton  particle: $\nu_{\tau}$ and $ \bar{\nu_{\tau}}$. \\

\begin{figure}[]
\begin{center}
\includegraphics[width=0.8\textwidth]{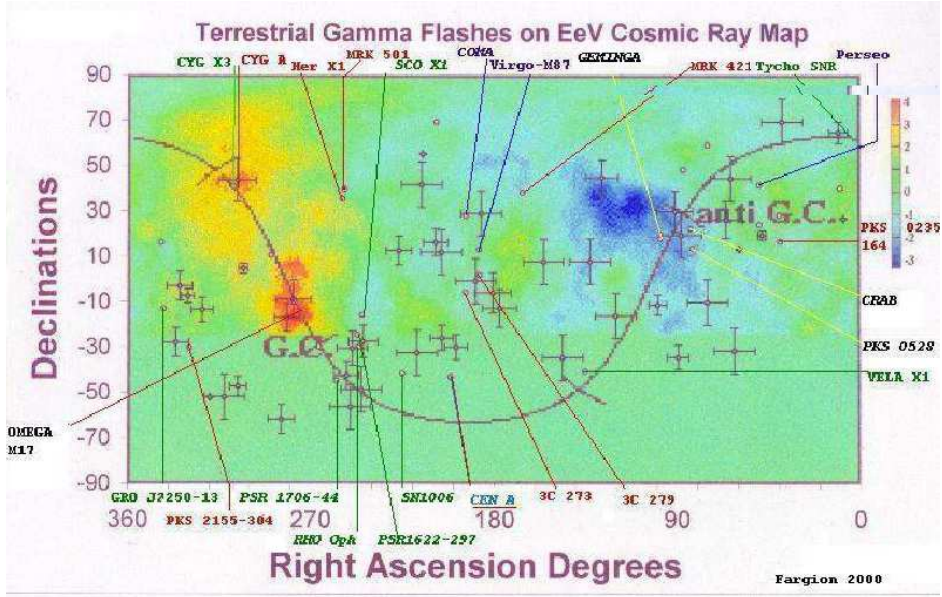}
\end{center}
\caption []{Terrestrial Gamma Flash in celestial coordinate over
UHECR diffused data by AGASA  cosmic rays at EeV energies.
Known powerful Galactic and Extra galactic sources, very active at X,$\gamma$,$TeV$,
possibly related to the TGF events are also shown by labels.  }
\label{fig:boxed_graphic 10a}
\end{figure}


\newpage



\end{document}